\documentclass[twocolumn,fleqn,natbib]{svjour2}
\bibpunct{[}{]}{;}{n}{}{,} 
\smartqed  
\usepackage{graphicx}
\usepackage{amsmath}
\newcommand{\pd}[2]{\frac{\partial {#1}}{\partial {#2}}}

\newcommand{\grad}{\vec{\nabla}}

\journalname{Granular Matter}
\begin{document}
\title{Plug Conveying in a Horizontal Tube}

\author{Martin Strau\ss \and Sean McNamara \and Hans J.\ Herrmann}

\institute{M. Strau\ss \at
Institut f\"ur Computerphysik, Universit\"at Stuttgart,\\
 70569 Stuttgart, GERMANY\\
              Tel.: +49-(0)711/685-3593\\
              Fax: +49-(0)711/685-3658\\
              \email{mys@ica1.uni-stuttgart.de}           
}

\date{Received: date }

\maketitle

\begin{abstract}
Plug conveying along a horizontal tube has been investigated through simulation,
using a discrete element simulation approach for the granulate particles
and a pressure field approach for the gas.
The result is compared with the simulation of the vertical plug conveying.
The dynamics of a slug are described by porosity,
velocity and force profiles. Their dependence on simulation parameters
provides an overall picture of slug conveying.
\end{abstract}

%
%
\section{Introduction}

A quite common method for the transportation of 
granular media is pneumatic conveying, where grains are driven
through pipes by air flow.
Practical applications for pneumatic conveying can be found in food industry
and in civil and chemical engineering.
One distinguishes two modes of pneumatic conveying: dilute and dense phase
conveying.
Dilute phase conveying has been studied in much detail%
~\cite{ALA0126,DAS0130,RAU0121,VAN0062,MAS0056,BIL0153,YAM0039,MAS0138}
and is well understood.
The grains are dispersed and dragged individually by
the gas flow and the interaction between grains is weak.
This is not true for dense phase conveying,
where the particle interaction is important
and where particle density waves can be observed.
The behavior of these density waves depends strongly on the orientation
of the tube.
In vertical plug conveying, most of the particles are located in ``plugs''
(dense regions) that are pushed up the tube by the 
pressure gradient~\cite{STR0192}.
In a horizontal tube the granular medium separates into two
layers, the slowly moving bed at the bottom of the pipe and
the traveling ripples or the ``slugs''%
~\cite{TSU0093,TOM0147,MAS0154,ZHU0120,VAS0136},
where slugs are accumulations of particles which fill up the cross-section
of the tube and move quickly in the direction of transportation.
Slug conveying occurs when the ratio of grain flux to gas flux is high.
This transport mode is sometimes also called ``plug conveying'',
in this paper this term is used in reference to vertical conveying.
Currently plug conveying is gaining importance in industry, because
it causes less product degradation 
and pipeline erosion than dilute phase conveying.

Unfortunately, current models~\cite{KON0095,SIE0096}
of slug conveying disagree
even on the prediction of such basic quantities
as the pressure drop and the total mass flow, and
these quantities have a great impact in the industrial application.
One of the reasons for the lack of valid models is that
it is difficult to study slugs experimentally in a detailed way.
Usually experimental setups are limited to the measurement of the
local pressure drop, the total mass flux and the velocity of the slugs.
The most promising experimental studies have been performed by electrical capacity
tomography~\cite{JAW0104,ZHU0120,ARK0105} and 
stress detectors~\cite{NIE0111,NIE0112,VAS0136}.
Plenty of experimental research has been done on the prediction of the
total pressure drop along a horizontal conveying line%
~\cite{GUI0167,PAN0184,MUS0186,PAN0139,MAS0154,LAO0134}
Simulational studies are handicapped by the high computational cost
for solving the gas flow and the particle-particle interaction,
and are therefore mostly limited to two dimensions.
For the dense phase regime, simulations have been done for 
bubbling fluidized beds~\cite{YEM0122,TSU0092,KAW0089,HUI0061,GOL0125,YUU0159}, which show at high gas velocities the
first signs of pneumatic transport~\cite{LIM0115,XUB0118,HOO0119},
for the strand type of conveying~\cite{HEL0059,TSU0058,TSU0091,YON0098},
and for slug conveying.
The simulation results on slugs published
by Tsuji et al~\cite{TSU0093}, Tomita et al~\cite{TOM0147}
and Levy~\cite{LEV0155} are discussed later in this paper.

The goal of this paper is to provide a detailed view of slugs, by using
a discrete element simulation combined with a solver for the pressure drop.
This approach provides access to important parameters like the porosity
and velocity of the granulate and the shear stress on the wall
at relatively low computational cost.
Contrary to the experiments,
it is possible to access these parameters 
with high spatial resolution and without influencing the process of 
transportation. 
Additionally to slug profiles, characteristic curves of the pressure drop 
and the influence of simulation parameters are measured.
The simulation results are compared to the results for
vertical plug conveying \cite{STR0192}.
%
%
%
%
\section{Simulation Model}
Plug conveying is a special case of the two phase flow of grains and
gas. It is therefore necessary to calculate the motion of both phases,
as well as the interaction between them.
In the following, we explain how our algorithm treats each of these problems.
\subsection{Gas Algorithm}
The model for the gas simulation was first introduced by
McNamara and Flekk{\o}y \cite{MCN0003} and has been implemented for
the two dimensional case
to simulate the rising of bubbles within a fluidized bed.
For the simulation of slug conveying we developed a three dimensional version
of this algorithm.

The algorithm is based on the mass conservation of the gas
and the granular medium.
Conservation of grains implies that
the density $\rho_p$ of the granular medium obeys
\begin{equation}
\pd{\rho_p}{t}+\vec{\nabla}\cdot(\vec{u}\rho_p)=0,\qquad \rho_p=\rho_{s}(1-\phi),
\end{equation}
where the specific density of the particle material is $\rho_{s}$,
the porosity of the medium is $\phi$
(i.e. the fraction of the space available to the gas),
and the velocity of the granulate is $\vec{u}$.

The mass conservation equation for the gas is
\begin{equation}
\pd{\rho_g}{t}+\vec{\nabla}\cdot(\vec{v}_g\rho_g)=0,
\end{equation}
where $\rho_g$ is the density of the gas and $\vec{v}_g$ its velocity.
This equation
can be transformed into a differential equation for the gas pressure $P$
using the ideal gas equation $\rho_g\propto \phi P$, together with the assumption of uniform
temperature.

For small Reynolds numbers the velocity $\vec{v}_g$ of the gas is related to
the granulate velocity $\vec{u}$ through the d'Arcy relation:
\begin{equation}\label{equ:darcy}
-\vec{\nabla}P=\frac{\eta}{\kappa(\phi)}\phi(\vec{v}_g-\vec{u}),
\end{equation}
where $\eta$ is the dynamic viscosity of the air and $\kappa$ is the
permeability of the granular medium. This relation was first given
by d'Arcy in 1856 \cite{DAR0094}.
For the permeability $\kappa$ the Carman-Kozeny relation \cite{CAR0084}
was chosen,
which provides a relation between the porosity $\phi$, the particle diameter $d$
and the permeability of a granular medium of monodisperse spheres,
\begin{equation}
\kappa(\phi)=\frac{d^2\phi^3}{180(1-\phi)^2}.
\end{equation}
After linearizing around
the normal atmospheric pressure $P_0$ the resulting differential equation
only depends on the relative pressure $P^\prime$ ($P=P_0+P^\prime$),
the porosity $\phi$ and the granular velocity $\vec{u}$,
which can be obtained from the particle simulation,
and some material constants like the viscosity~$\eta$:
\begin{equation}\label{eqn:dgl}
\pd{P^\prime}{t}=\frac{P_0}{\eta\phi}\vec{\nabla}(\kappa(\phi)\vec{\nabla}P^\prime)-\frac{P_0}{\phi}\vec{\nabla}\vec{u}.
\end{equation}
This differential equation can be interpreted as a diffusion equation with
a diffusion constant $D=\phi\kappa(\phi)/\eta$.
The equation is solved numerically, using a
Crank-Nickelson approach for the discretization. Each dimension is integrated
separately.

The boundary conditions are imposed by
adding a term $\mp S$ on the right hand side of equation (\ref{eqn:dgl})
at the top and the bottom of the tube, where $S\propto v_gP_0$.
This mimics a constant gas flux with velocity $v_g$
at a pressure $P_0$ into and out of the tube.
\subsection{Granulate Algorithm}
The model for the granular medium simulates each grain
individually using a discrete element simulation (DES).
For the implementation of the discrete element simulation we used 
a version of the molecular dynamics method described by 
Cundall and Strack~\cite{CUN0172}.
The particles are approximated by monodisperse spheres and rotations in
three dimensions are taken into account.

The equation of motion for an individual particle is
\begin{equation}
m\ddot{\vec{x}}=m\vec{g}+\vec{F}_{c}-\frac{\grad P}{\rho_s(1-\phi)},
\end{equation}
where $m$ is the mass of a particle, $\vec{g}$ the gravitation constant and
$\vec{F}_{c}$ the sum over all contact forces. The last term,
the drag force, is assumed to be a volume force
given by the pressure drop $\grad P$ and the local mass density of
the granular medium $\rho_s(1-\phi)$.

The interaction between two particles in contact is given
by two force components:
a normal and a tangential component with respect to the particle surface.
The normal force is the sum of a repulsive elastic force (Hooke's law)
and a viscous damping.
The tangential force is proportional to the minimum of the normal force
(sliding Coulomb friction) and a viscous damping.
The viscous damping is used when the relative
surface velocity of the particles in contact is small.
The same force laws are considered for the interaction between particles 
and the tube wall.
\subsection{Gas-Grain Interaction}
The simulation method uses both a continuum and a
discrete element approach. While the gas algorithm
uses fields, which are discretized on a cubic grid,
the granulate algorithm describes discrete particles in a continuum.
An interpolation is needed for the algorithms to interact.
For the interpolation a tent function $F(\vec{x})$ is used:
\begin{equation}
F(\vec{x})=f(x)f(y)f(z),\qquad f(x)=\begin{cases}
  1-|x/l|, &  |x/l| \le 1, \\
  0, &  1<|x/l|,
\end{cases}
\end{equation}
where $l$ is the grid size used for the discretization
of the gas simulation.

For the gas algorithm the porosity $\phi_j$ 
and the granular velocity $\vec{u}_j$
must be derived from the particle positions $x_i$ and velocities $v_i$,
where $i$ is the particle index and
$j$ is the index of the grid node.
The tent function distributes the particle properties around
the particle position smoothly on the grid:
\begin{equation}
\phi_j=1-\sum_i F(\vec{x}_i-\vec{x}_j), \qquad
\vec{u}_j=\frac{1}{1-\phi_j}\sum_i \vec{v}_i F(\vec{x}_i-\vec{x}_j),
\end{equation}
where $x_j$ is the position of the grid point 
and the sum is taken over all particles.

For the computation of the drag force on a particle
the pressure drop $\grad P_i$ and the
porosity $\phi_i$ at the position of the particle are needed.
These can be obtained by a linear interpolation of the fields $\grad P_j$
and $\phi_j$ from the gas algorithm:
\begin{equation}
\phi_i=\sum_j \phi_j F(\vec{x}_j-\vec{x}_i), \qquad 
\grad P_i=\sum_j \grad P_j F(\vec{x}_j-\vec{x}_i),
\end{equation}
where the sum is taken over all grid points.
%
%
%
%
\section{Simulational Results}
The setup for the simulation consists of a horizontal tube of length $L_t=52.5\,cm$ 
and of internal diameter $D_t=7mm$.
The air and the granular medium is injected at a constant mass flow rate 
at on end of the tube.
At the beginning of a simulation the tube is empty.
The default parameters were chosen 
to be the same as in an earlier paper~\cite{STR0192} 
for vertical plug conveying.
As default the mass flow rate of the granular medium is $2.49\,kg/h$.
Default parameters for the particles are:
diameter $d=1.4\,mm$, density $\rho_s=937\,kg/m^3$,
Coulomb coefficient $\mu=0.5$ and restitution coefficient $e=0.5$.
The gas volume has been discretized into 150x2x2 grid nodes,
which corresponds to a grid constant of $3.5\,cm$.
The gas pressure is set to $P_0=1013.25\,hPa$, Simulations are preformed
for gas viscosities $\eta$ from $0.045\,cP$ to $0.085\,cP$
and gas flows $\dot{V}$ between $1.1\,l/min$ and $9.2\,l/min$.
The gas flow is usually given by the 
superficial gas velocity~$v_s=\phi v_g=4\dot{V}/\pi D_t^2$~\cite{HON0128},
where $v_g$ is the equivalent gas velocity for an empty tube.

\begin{figure*}[h]
\centering
\includegraphics[width=0.85\textwidth]{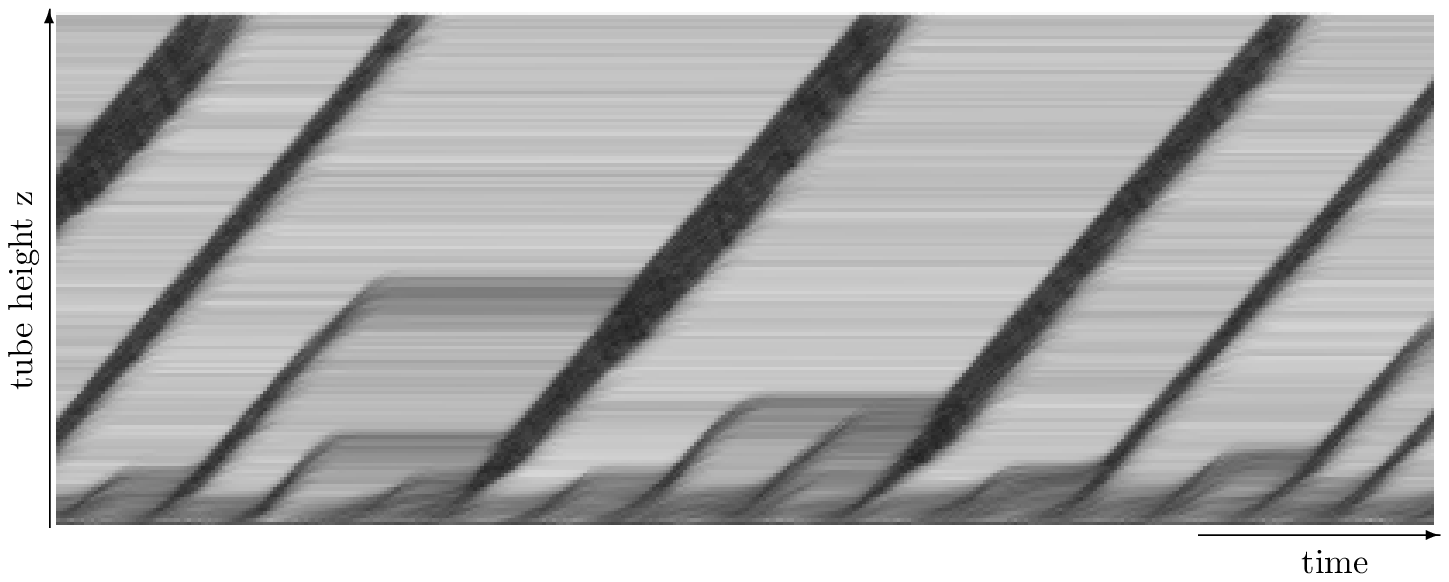}
\caption{%
Spatio-temporal image of the porosity along the horizontal tube.
The dark regions correspond to low porosity, light regions to high
porosity.
The entire tube ($52.5\,cm$) is displayed, the elapsed time is~$4\,s$.
Default parameters for the simulation were used; the superficial gas velocity
is $1\,m/s$, the gas viscosity is $0.0673\,cP$,
and the Coulomb coefficient is $0.5$.
Particles are introduced at the bottom. On average 3200 particles
are inside the tube, and the pressure drop is $10\,hPa/m$.
}
\label{fig:rho-W1020}%
\end{figure*}

\begin{figure*}[h]
\centering
\includegraphics[width=0.85\textwidth]{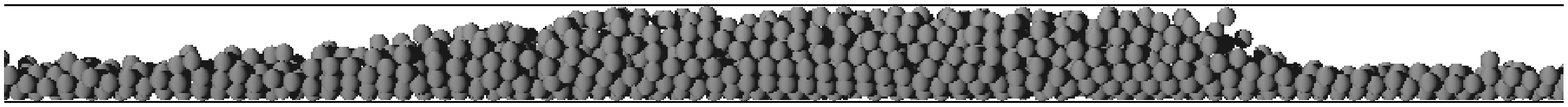}
\caption{%
Image of a slug corresponding to the slug conveying
shown in figure~\ref{fig:rho-W1020}.
The direction of motion is from left to right with a superficial
gas velocity of $1\,m/s$. A stretch of $11.3\,cm$ length is displayed.
The slug contains about 470 particles with diameter $1.4\,mm$.
}
\label{fig:slug-W1020}%
\end{figure*}

The flow in the corresponding experiment for 
the vertical conveying~\cite{STR0192}
is turbulent (particle $Re\approx 65$).
In the simulation, an effective gas viscosity is used to account
for the effect of turbulence.
For an effective gas viscosity $\eta=0.0673\,cP$, 
a gas flow $\dot{V}=2.3\,l/min$
and a Coulomb coefficient $\mu=0.5$
slug conveying is observed as shown in figure~\ref{fig:rho-W1020}.
The measured pressure drop is $10\,hPa/m$, the slug velocity is $0.42\,m/s$,
and the slug length is 4-8$\,cm$.
An image of a corresponding slug is shown in figure~\ref{fig:slug-W1020}.

As pointed out in the introduction, some simulations of 
horizontal slug conveying have already been published.
The horizontal transport of a slug has been studied by Tsuji et al~\cite{KAW0090}.
The tube with diameter $5\,cm$ and length $80\,cm$ contained 1000 particles
with a diameter of 1\,cm, which the same diameter ratio as used in this study.
Contrary to our simulations, the slug was created
by the initial conditions, where a given range of the tube was filled with particles.
Nevertheless the qualitative behavior of the particles is the same
as shown in figure~\ref{fig:slug-W1020} and \ref{fig:slug-phi}.

Tomita et al~\cite{TOM0147} assumed the slugs to be indivisible objects.
The computed trajectories of the slugs are similar to the trajectories
observed in our studies, however he neglects the possibility of slugs
to grow or even to dissolve which is seen in our results.

Levy~\cite{LEV0155} applied a two fluid approach on the horizontal conveying.
He observes the break up of a large artificially build slug into smaller
slug.
Such behavior is observed neither in our simulation nor in Tsuji's.

\subsection{Characteristic curves}
\begin{figure}[h]
\centering
\includegraphics[width=0.85\columnwidth]{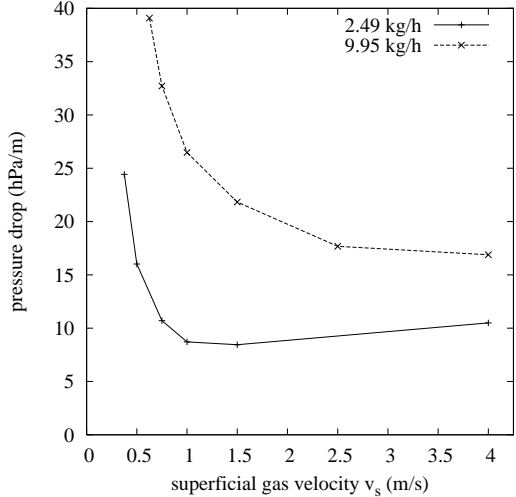}
\caption{%
Total pressure drop against superficial gas velocity for different
granular mass flows. Plotted are characteristic curves
from the simulation for the granular mass flows $2.49\,kg/h$ and $9.95\,kg/h$.
For mass flow $2.49\,kg/h$ bulk transport is observed when
the superficial gas velocity is below $0.25\,m/s$.
}
\label{fig:v-dp}%
\end{figure}

The ``characteristic curves'' of a pneumatic transport system
are plots of the pressure drop against the superficial gas velocity
$v_s=\phi v_g$ for different mass flows of the granulate.
This kind of diagram is
highly dependent on the material characteristics of the tube wall and
the granulate and can be used to predict the overall transport performance
for given parameter sets.
Such a diagram, from data of our simulation,
is shown in figure~\ref{fig:v-dp}.

The diagram provides the typical qualitative behavior for pneumatic transport.
From top left to bottom right with increasing superficial gas velocity,
three regions can be distinguished.
First, for small superficial gas velocities~($v_s<0.38\,m/s$),
one has bulk transport.
The tube is completely filled with granulate, so the pressure drop is high.
Nevertheless the drag force on the bulk is 
too small to overcome the friction between the granular medium and the wall. 
In this case the transport comes through the enforced granular mass flow at the inlet of the tube.

For moderate velocities ($0.38\,m/s\le v_s<4\,m/s$), slug conveying is observed
(fig.~\ref{fig:rho-W1020}).
The particles injected at the inlet organize into slugs.
After a short acceleration, the slugs move forward
with a constant velocity.
A slug always leaves particles behind it
and usually maintains its length by collecting particles in front of it.
A slug disintegrates when it gets too small.
The particles behind the slug rest at the bottom of the tube.
The amount of particles left behind a slug is independent of the slug size.
Larger particle amounts left by a dissolved slug 
are collected by the following slug and cause it to grow.
The porosity of the granular medium inside a slug
is close to the minimum porosity, and the slug edges are smooth.

\begin{figure*}[h]
\centering
\includegraphics[width=0.85\textwidth]{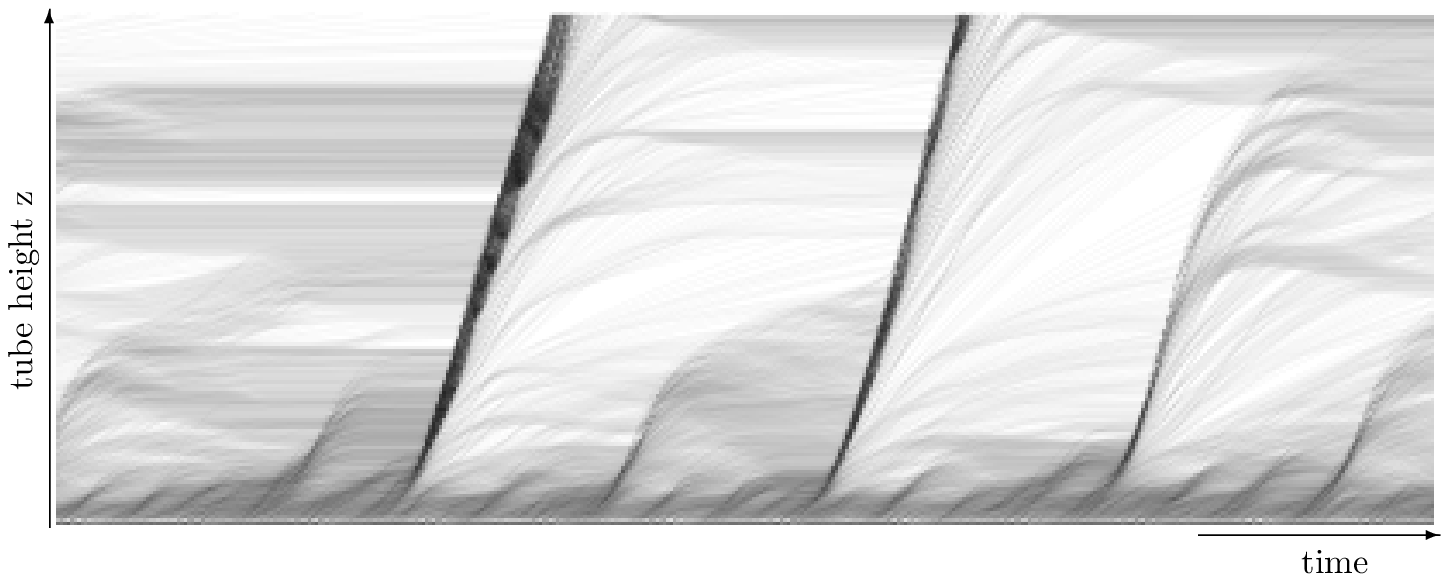}
\caption{%
Spatio-temporal image of the porosity along the horizontal tube.
The dark regions correspond to low porosity, light regions to high
porosity.
The entire tube ($52.5\,cm$) is displayed, the elapsed time is~$4\,s$.
A higher superficial gas velocity (4\,m/s)
has been used compared to the spatio-temporal image in figure~\ref{fig:rho-W1020}
($1\,m/s$).
On average 1400 particles are within the tube, and the pressure drop is $10\,hPa/m$.
}
\label{fig:rho-W1003}%
\end{figure*}

For high superficial gas velocities~($v_s>4\,m/s$) the tube is almost empty
(fig.~\ref{fig:rho-W1003});
in the simulation the particles are pushed out either as small slugs
or as individual particles sliding on the bottom of the tube.
The porosity increases with the superficial gas velocity.
In this region the simulation method underestimates the pressure drop,
because it does not consider the increasing drag force on single particles,
which in the experiment dominates in this region.
The boundaries between the described regions depend on the simulation parameters.

A nearly proportional relation is observed
between the total pressure drop and the total number of particles in the tube.
This can be explained by the observation that
the amount of particles between the slugs is small and
in the slugs most particles are densely packed at a well 
defined porosity.
Through d'Arcy's law,
the total pressure drop depends linearly on the tube length
filled with this porosity, since the pressure drop on the granulate between the
slugs is negligible and causes only a small deviation.
At high gas velocities this is no longer true, because 
the plugs no longer contain the majority of the particles.

In the following, the dependence of the pressure drop on 
the mass flux of the granulate,
the air viscosity $\eta$, the atmospheric pressure $P_0$ and the
Coulomb coefficient $\mu$ is discussed.
For the parameter studies the superficial gas velocity has been
fixed to $1\,m/s$. 
For higher velocities the sensitivity to the parameter values decreases.

\begin{figure}[h]
\centering
\includegraphics[width=0.85\columnwidth]{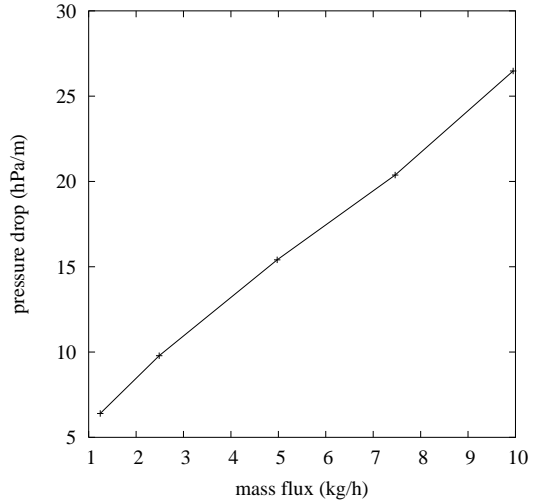}
\caption{%
Dependence of the pressure drop on the mass flux of the granulate
at a superficial gas velocity $1\,m/s$.
}
\label{fig:m-dp}%
\end{figure}
As one can see in figure~\ref{fig:m-dp} the pressure drop increases 
linearly with mass flux of the granulate. The increase in the
pressure drop is associated with
an increase of the number of slugs within the tube.

The gas flow can be influenced by changing the atmospheric pressure $P_0$
or the diffusion constant $D$.
An increase in background pressure $P_0$ combined with an increase
in superficial velocity leaves the pressure drop unchanged.
This can be deduced directly from equation (\ref{eqn:dgl})
by noting that both terms on the right hand side are proportional to $P_0$.

\begin{figure}[h]
\centering
\includegraphics[width=0.85\columnwidth]{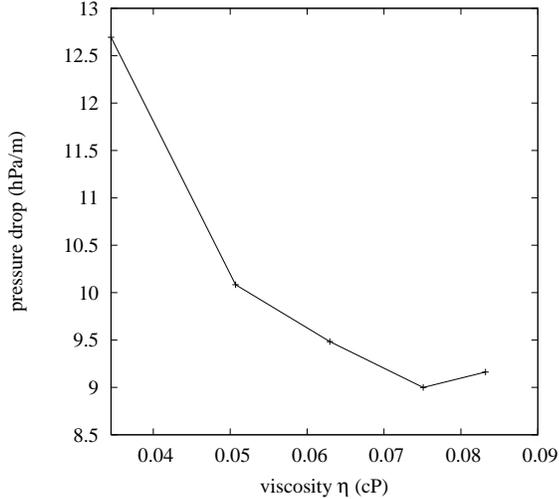}
\caption{%
Dependence of the pressure drop on the dynamic viscosity $\eta$
of the gas at a superficial gas velocity $1\,m/s$.
}
\label{fig:vis-dp}%
\end{figure}

The diffusion constant $D\propto d^2/\eta$ can be changed
through the particle diameter $d$ and the viscosity $\eta$.
Therefore it is sufficient to analyze the parameter space for the viscosity
at a constant diameter as shown in figure~\ref{fig:vis-dp}.
The pressure drop is decreasing with increasing viscosity.

The parameter of the particle simulation with most 
influence on the transport of the granular medium is the Coulomb coefficient
$\mu$.
The restitution coefficient has only a small effect,
except when it is unrealistically large.

\begin{figure}[h]
\centering
\includegraphics[width=0.85\columnwidth]{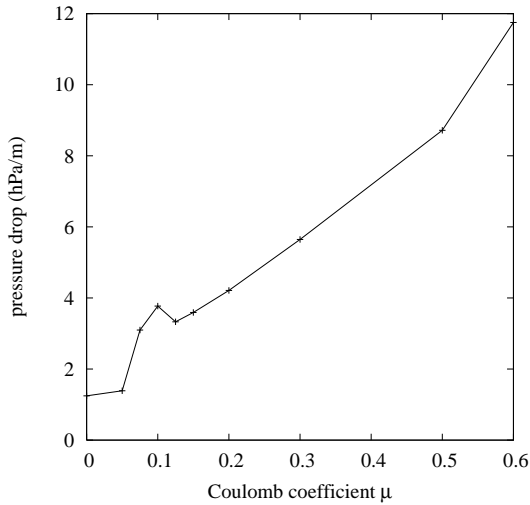}
\caption{%
Dependence of the pressure drop on the Coulomb coefficient $\mu$.
The superficial gas velocity is $1\,m/s$.
For low Coulomb coefficients ($\mu<0.1$) even small particle groups 
are able to slide within the tube, so slugs do not occur.
}
\label{fig:cc-dp}%
\end{figure}
As one can see in figure~\ref{fig:cc-dp} the pressure drop increases with
$\mu$.
For low Coulomb coefficients ($\mu<0.1$) the injected particles are
sliding individually along the tube with accelerating speed.
As most of the tube is empty, the total pressure drop
is small.
As $\mu$ increases, the particles start first to
slide as one layer.
Then the particle layer gets slower and thicker
and therefore causes an higher pressure drop.
At a Coulomb coefficient of $\mu=0.1$, a peak in the pressure drop 
coincides with the transition from the conveying by a sliding particle layer
to slug conveying.
Slugs first occur when the cross-section of the tube is filled locally
somewhere along the tube.
They are more efficient in transporting particles, therefore
the pressure drop first decreases until a pure slug conveying
is reached ($0.1<\mu<0.15$).
For higher coefficients ($\mu>0.15$) the slugs become slower
due to the growing friction and cause therefore
a rising pressure drop.

\subsection{Slug statistics}

The spatio-temporal image of the porosity along the horizontal tube
(fig.~\ref{fig:rho-W1020}) provides a rough picture of slugs and their
movement along the tube. A statistical approach is necessary to get
more precise values.
Properties of interest are the porosity and the granular velocity within a slug,
the slug length, and their dependence on the horizontal position $x$ 
of the slug within the tube.
To get some average values for the porosity and the granular velocity,
the tube was segmented into horizontal slices of length $3.5\,mm$.
For each slice the average porosity and granular velocity was computed
every $0.01\,s$.
The contribution of a particle was weighed by the volume occupied
by that particle within a given slice.

The resulting vertical porosity was used to identify slugs.
Every region with a porosity lower than 0.6 is defined to belong to a slug.
\begin{figure}[h]
\centering
\includegraphics[width=0.85\columnwidth]{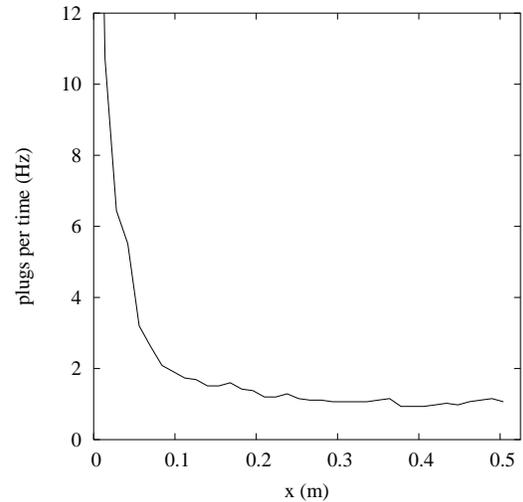}
\caption{%
Number of slugs per time as a function of tube position $x$.
The corresponding spatio-temporal image is shown in figure \ref{fig:rho-W1020}.
The total tube length is $52.5\,cm$, and default parameters are used.
The data is averaged over $22.5\,s$.
}
\label{fig:W1020_slug_number_rate_versus_x}%
\end{figure}

Figure \ref{fig:W1020_slug_number_rate_versus_x} shows the 
slug rate, or number of slugs 
per time as a function of the position along the tube.
At the inlet of the tube the incoming granular medium fragments into many
small slugs. Even though their velocity is low, the resulting
slug rate is high.
As can be seen in figure~\ref{fig:rho-W1020}, many slugs dissolve
along the tube, therefore the slug rate decreases.
Most slugs dissolve within the first $0.3\,m$ along the tube.
Beyond $x=0.3\,m$ the number of slugs remains stable 
until they leave the system.

For each slug, the center of mass, the minimal porosity,
the maximal granular velocity and the slug length have been computed.

\begin{figure}[h]
\centering
\includegraphics[width=0.85\columnwidth]{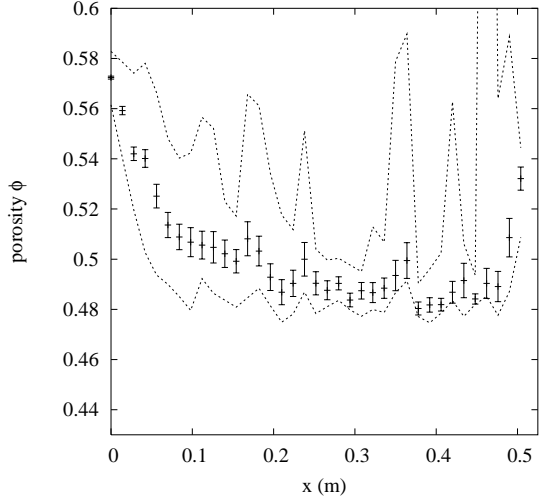}
\caption{%
Minimal porosity in slugs at tube position $x$,
corresponding to the slug rate shown in figure \ref{fig:W1020_slug_number_rate_versus_x}.
The bars denote the uncertainty of the average, and the dotted lines indicate
the width of the distribution;
at each value of $x$, half of the observed slugs have a porosity 
lying between the upper and the lower dotted lines.
The slug porosity decreases when the granulate enters the tube
($x<0.2\,m$) and increases when it leaves the tube ($x>0.45\,m$).
}
\label{fig:W1020_slug_phi_versus_x}%
\end{figure}

\begin{figure}[h]
\centering
\includegraphics[width=0.85\columnwidth]{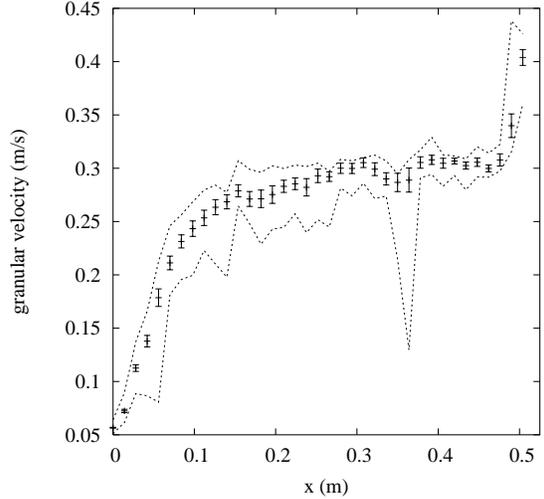}
\caption{%
Maximal granular velocity in slugs at tube position $x$,
corresponding to the slug rate shown in figure \ref{fig:W1020_slug_number_rate_versus_x}.
The bars and dotted lines have the same meaning as in 
figure~\ref{fig:W1020_slug_phi_versus_x}.
The granular velocity saturates in the middle of the tube, the increase
of the granular velocity at the beginning and at the end of the tube
is due to the boundary effects described in figure \ref{fig:W1020_slug_phi_versus_x}.
}
\label{fig:W1020_slug_uxx_versus_x}%
\end{figure}

\begin{figure}[h]
\centering
\includegraphics[width=0.85\columnwidth]{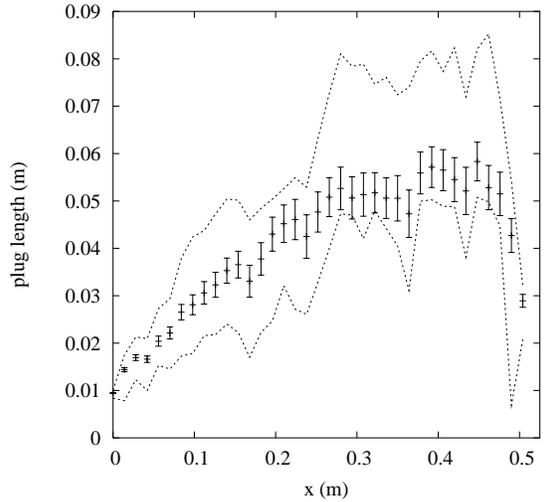}
\caption{%
Mean slug length at tube position $x$,
corresponding to the slug rate shown in figure \ref{fig:W1020_slug_number_rate_versus_x}.
The bars and dotted lines have the same meaning as in 
figure~\ref{fig:W1020_slug_phi_versus_x}.
The average slug length increases along the tube, consistent with the
decrease in the slug rate (fig.~\ref{fig:W1020_slug_number_rate_versus_x}).
}
\label{fig:W1020_slug_length_versus_x}%
\end{figure}

Figure \ref{fig:W1020_slug_phi_versus_x} shows 
the minimum slug porosity as a function of the 
horizontal position $x$ of a slug. 
At each value of $x$, the mean porosity and its uncertainty 
(standard deviation divided by the square root of the number of slugs)
were calculated.
These quantities are shown by the bars in figure~\ref{fig:W1020_slug_phi_versus_x}.
To show the distribution of porosity about the mean,
the two dotted lines were added.
At each position $x$, half of the slugs have a porosity lying between 
these two lines. 
The same analysis was carried out for the data in Figures \ref{fig:W1020_slug_uxx_versus_x} and \ref{fig:W1020_slug_length_versus_x}.
As one can see in figure~\ref{fig:W1020_slug_phi_versus_x}
at the left of the graph,
the granular medium is inserted at the inlet
of the tube with a porosity of 0.57.
From there the porosity decreases quickly to about 0.47 at a height $x=0.2\,m$
and then remains almost constant until $x=0.45\,m$.
At the end of the tube ($x>0.45\,m$) the porosity increases again until the
grains leave the simulation space at $x=0.525\,m$.

Figure \ref{fig:W1020_slug_uxx_versus_x} shows the corresponding particle velocity.
The change in porosity comes along with an increase of the granular velocity
within the slugs.
The granulate is inserted with an initial velocity of $0.04\,m/s$.
The granular velocity saturates to a final
velocity ($0.3\,m/s$) at a position of $0.2\,m$.
Starting at about $0.45\,m$, the granular medium accelerates
until the grains leave the tube.

An explanation for the constant slug velocity in the middle of the tube
can be derived
using the balance equation for the forces on a slug:
\begin{equation}
 F=-F_c+\alpha(\phi)(v_g-u).
\end{equation}
where 
$F_c$ is the force on the slug due to the friction with the wall.
The last term 
$\alpha(\phi)(v_g-u)$ is the drag force on the slug, which is proportional
to the relative velocity between the particles and the gas 
within the slug.
At the middle of the tube the porosity of the slug $\phi$ is constant.
On small time scales 
the gas velocity $v_g$ can be assumed constant.
Only the drag force depends on the granular velocity $u$ through
equation~(\ref{equ:darcy}).
For a certain granular velocity the drag force balances
the friction forces, and thus $u$ remains constant.
The solution is stable under small fluctuations in $u$.

Figure \ref{fig:W1020_slug_length_versus_x} shows the mean slug length along the tube.
The average slug length increases along the tube, due to the collection
of particles of dissolved slugs by the following ones.
Beyond $x=0.3\,m$ a stable slug length is reached.
The increase of the slug length combined with
the decrease of the number of slugs per time (fig.~\ref{fig:W1020_slug_number_rate_versus_x})
conserves the mass flux of the granulate. 

As can one seen from Figs. \ref{fig:W1020_slug_phi_versus_x},
\ref{fig:W1020_slug_uxx_versus_x} and \ref{fig:W1020_slug_length_versus_x},
the boundary condition at $x=0.525\,m$ affects only a small region close to the end of the tube.
In vertical plug conveying, this boundary region is much larger.
This difference arises because the grains between the plugs 
or slugs have different behaviors.
In vertical conveying, particles are falling downwards,
so that when a plug is removed at the top of the tube,
the flux of particles onto the next plug is reduced shortly thereafter.
On the other hand, in horizontal conveying, the grains between the slugs are
simply lying on the bottom of the tube. When a slug is removed at the end
of the tube, this information is not transmitted to the next plug.
The succeeding slug is only influenced by the thickness of 
the particle layer left behind.
Another consequence of this difference is that,
contrary to the vertical conveying,
the distance between slugs has no
effect on their interaction.

The slug profiles (Fig.~\ref{fig:rho_x}-\ref{fig:eneryyz_x}) indicate
that the slugs only interact through the thickness of the particle layer
left behind them.
Defining everything with a porosity lower then 0.6 as belonging to a slug,
the interaction range with the particle layer is of the size of the slug length.

The diagrams in figures~\ref{fig:W1020_slug_phi_versus_x} to \ref{fig:W1020_slug_length_versus_x} imply that
there is a typical porosity, granular velocity and length
of slugs and a characteristic slug profile 
for a given position along the tube exists.
In the following averaged vertical and radial profiles of slugs
at the position of $0.26\,m$ are discussed.
To get some sensible profiles, slugs 
with length and granular velocity close to the mean values
($L_p=0.05\pm0.03\,m$, $u=0.29\pm0.1\,m/s$) were selected.

\subsection{Horizontal slug profiles}
While in experiments recognizing and measuring parameters 
for global slug conveying are rather simple,
the measurement of profiles for individual slugs remains
a nearly impossible task. So one of the reasons for simulating slug conveying
is to provide a detailed picture of what happens within a slug, and how
parameters like porosity, granular velocity, or shear stress change along
the slug.

\begin{figure}[h]
\centering
\includegraphics[width=0.85\columnwidth]{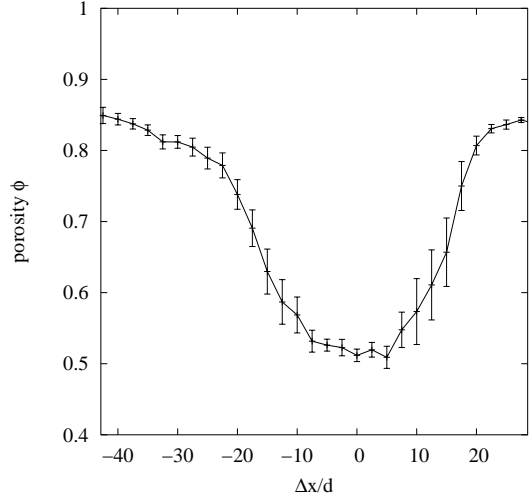}
\caption{%
Horizontal porosity profile along an averaged slug,
containing about 470 particles,
positioned at the middle of the tube. The profile was averaged over 
six slugs with
granular velocity $0.29\pm0.1\,m/s$ and slug length $0.05\pm0.03\,m$.
The data belong to the simulation displayed in figure~\ref{fig:rho-W1020}.
The horizontal axis denotes the relative horizontal position $\Delta x$ 
along the tube with respect to the center of mass. The horizontal position
is given in multiples of the particle diameter $d=1.4\,mm$.
}
\label{fig:rho_x}%
\end{figure}

\begin{figure}[h]
\centering
\includegraphics[width=0.85\columnwidth]{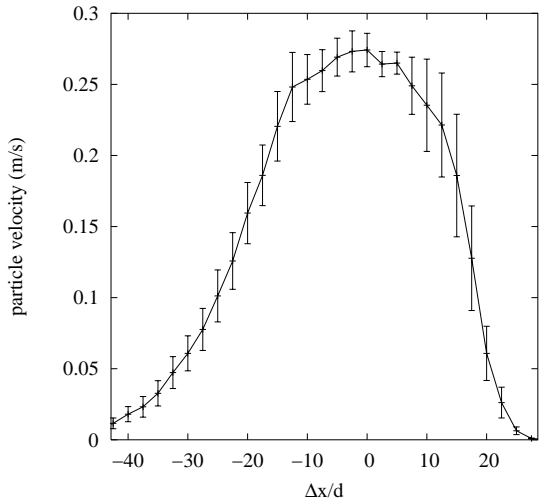}
\caption{%
Averaged velocity of the granular medium along 
the averaged slug of figure~\ref{fig:rho_x}.
Inside the region around the slug ($\Delta x/d<-40$ \& $\Delta x/d>30$) the 
granular medium is moving forward,
outside it is resting.
}
\label{fig:uxx_x}%
\end{figure}

\begin{figure}[h]
\centering
\includegraphics[width=0.85\columnwidth]{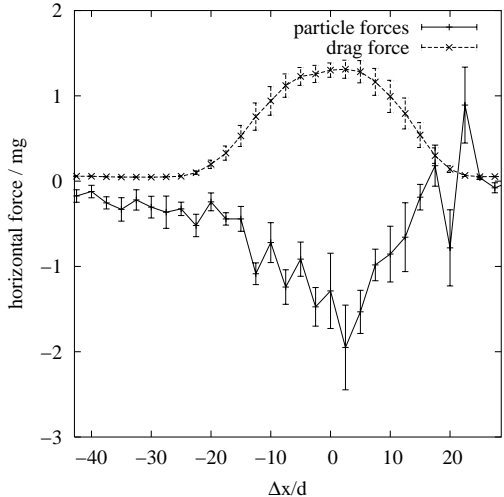}
\caption{%
Forces in the direction of motion acting on particles along the averaged slug 
in figure~\ref{fig:rho_x}.
Interparticle forces and friction are summed up and displayed
as particle force. The drag force corresponds to the pressure drop of the gas.
Outside of the slug ($\Delta x/d<-40$ \& $\Delta x/d>30$) the particle force is zero.
The fluctuations in the particle force ($x/d\approx20$) are strongest in
the collision region,
where the moving particles within the slug collide with the resting particles 
before the slug.
}
\label{fig:force_x}%
\end{figure}

\begin{figure}[h]
\centering
\includegraphics[width=0.85\columnwidth]{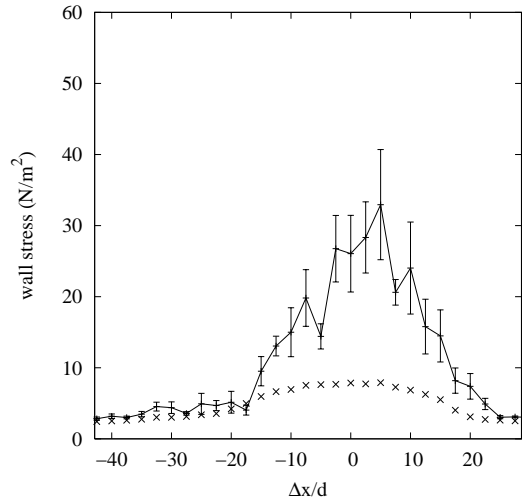}
\caption{%
Stress between the wall and the granular medium along the averaged slug
of figure~\ref{fig:rho_x}.
The low wall stress before and after the slug corresponds to the weight of
the layer of particles resting at the bottom of the tube,
which is given by the dotted curve.
}
\label{fig:wallstress_x}%
\end{figure}

\begin{figure}[h]
\centering
\includegraphics[width=0.85\columnwidth]{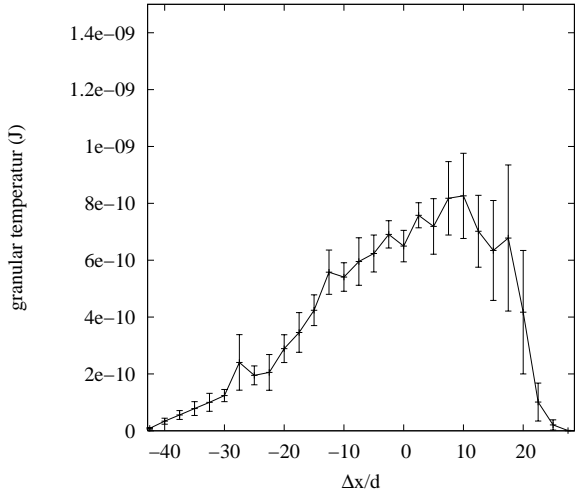}
\caption{%
Granular temperature along the averaged slug of figure~\ref{fig:rho_x}.
The granular temperature increases rapidly at the front of the slug.
Due to the large damping ($e=0.5$) the granular temperature is dissipated.
Behind the center of the slug the temperature decreases linearly
until it vanishes behind the slug.
}
\label{fig:eneryyz_x}%
\end{figure}

Figure \ref{fig:rho_x} to \ref{fig:eneryyz_x} present averages of different
quantities over six slugs.
These slugs were taken from the middle of the tube $x=0.26\,m$ with
granular velocity $0.29\pm0.1\,m/s$ and slug length $0.05\pm0.03\,m$.
The coordinate $\Delta x$ denotes the relative vertical position along the tube 
with respect to the center of mass.
The porosity profile of the slugs is shown in Figure \ref{fig:rho_x}.
At the front side of the slug, on the right hand side of figure \ref{fig:rho_x},
the porosity decreases from 85\% to $50\%$.
In the middle the porosity of the slug remains almost constant,
and increases at the backside of the slug.
The high porosity before and after the slug corresponds to a dense
particle layer at the bottom of the tube and a region devoid of particles
above this layer (fig.~\ref{fig:slug-phi}).

Figure \ref{fig:uxx_x} displays the velocity profile of
the granular medium for the slugs shown
in Figure \ref{fig:rho_x}. One can distinguish four different
regions: Ahead of the slug ($\Delta x/d\ge 30$),
the high porosity corresponds to the few particles resting at the bottom of the tube, averaged over the
cross-section of the tube.
These particles originate from the backside of a preceding slug.
The friction outweighs the drag force on the particles.
At the front side of the slug ($10\ge \Delta x/d>30$),
the particles accelerate, as they are pushed by the low porosity region.
In the following, this region at the front of the slug 
will be called collision region.
Inside the slug, where the porosity settles to a low value, the 
average granulate velocity is almost constant ($|\Delta x/d|>10$).
At the back side ($-10<\Delta x/d\le-40$) of the slug the granular medium
slows down until it rests again.
Thus the slug is always loosing material at the back side.
This region will be called disintegration region.

The trajectory of a single particle through the slug can be sketched
by a snapshot of the horizontal forces acting on it. Figure \ref{fig:force_x}
displays the drag force $F_d$ and the sum over the interparticle
and friction forces on a particle (here called particle forces) $F_p$.
These forces are averaged over the cross-section of the tube.
Before entering the slug,
the drag force on the particles is small, because the air is able to pass
above the particles.
The large fluctuations in the particle force arise 
when the  moving particles within the slug collide with the resting particles 
before the slug.
Within this region the particles are piled up,
until the cross-section of the tube is completely filled.
In the slug the drag force and the friction force balance each other.
Behind the slug, the drag force becomes negligible and
the particles are slowed down by the friction with the wall
until they come to rest.

The normal wall stress corresponding to the slug in Figure \ref{fig:rho_x}
is shown in Figure \ref{fig:wallstress_x}. In horizontal conveying
the normal wall stress is the sum of the normal interparticle forces
and the weight of the particles.
The low wall stress before and after the slug corresponds to the weight of
the layer of particles left behind by the slugs.
Higher wall stresses are produced by the slug including the collision 
and disintegration region. The wall stress increases
within the collision region and then decreases
in the disintegration region.
As one can see, the wall stresses are much higher than can be accounted for
by the weight of the material. Therefore, the particle in the slug must
have a high granular temperature.

Figure \ref{fig:eneryyz_x} shows the granular temperature along a slug.
The granular temperature is the average kinetic energy of the particles
minus the kinetic energy of the motion of their center of mass.
The granular temperature rises rapidly in the collision region.
Due to the high damping ($e=0.5$) these temperatures decrease 
along the slug and vanish behind the slug. 

\subsection{Cross-sections of a slug}
\begin{figure*}[h]
\centering
\includegraphics[width=0.85\textwidth]{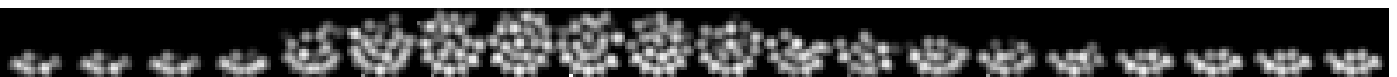}
\caption{%
Cross-sections of volume fraction of particles at half tube length.
The pictures were taken at a frame rate of $100\,Hz$.
From left to right an interval of $0.2\,s$ is displayed.
The greyscale from black to white corresponds to porosities from 100 to 0\%.
}
\label{fig:slug-phi}%
\end{figure*}

In horizontal slug conveying, the radial symmetry of the system
is broken by the gravitational force. The particles tend to segregate 
and settle on the bottom of the tube (fig.~\ref{fig:slug-phi}).
In the middle of the slug, the complete cross-section of the tube is
filled with particles. At the bottom of the tube, the particles
form layers along the wall.

\vspace*{2ex}
\begin{figure*}[h]
\centering
\includegraphics[width=0.85\textwidth]{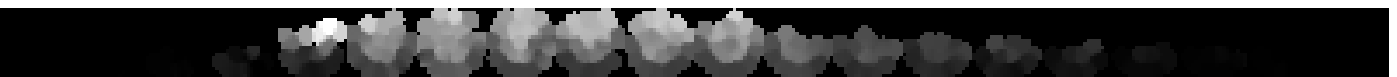}
\caption{%
Vertical velocity profile through the slug shown in figure~\ref{fig:slug-phi}
}
\label{fig:slug-uxx}%
\end{figure*}

Before the slug the particles are resting (fig.~\ref{fig:slug-uxx}).
At the front side of the slug the particles gain velocity in direction
of motion. This velocity decreases behind the slug until the particles
are again resting.

\begin{figure}[h]
\begin{center}
\includegraphics[width=0.85\columnwidth]{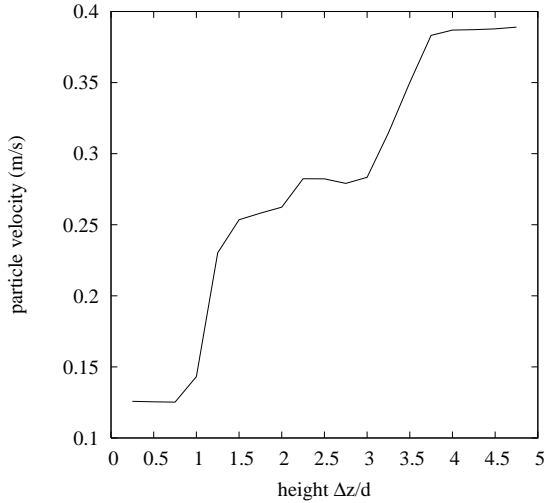}
\end{center}
\caption{\label{fig:uxx_z}%
Vertical velocity profile through the slug shown in figure~\ref{fig:slug-phi}
}
\end{figure}

Within the slug the lower particle layers are slower than the upper ones,
as shown in figure~\ref{fig:uxx_z}.
The particle layer at the bottom of the tube is slowed down due to the
high friction with the wall. 
The uppermost particles within the tube, moving freely between 
the upper tube wall
and the other particles below, have the highest velocity.
%
%
%
%
\section{Comparison of horizontal and vertical transport}
The same simulation method as used for horizontal conveying
has also been applied to vertical conveying.
A parameter study and a detailed analysis of the plugs for
the vertical plug conveying has been published separately \cite{STR0192},
with tube dimensions and default parameters for the particles and the gas
being identical.

\begin{figure}[h]
\centering
\begin{minipage}{0.85\columnwidth}
\hspace*{-4ex}(a)\\[-1ex]
\includegraphics[width=1\columnwidth]{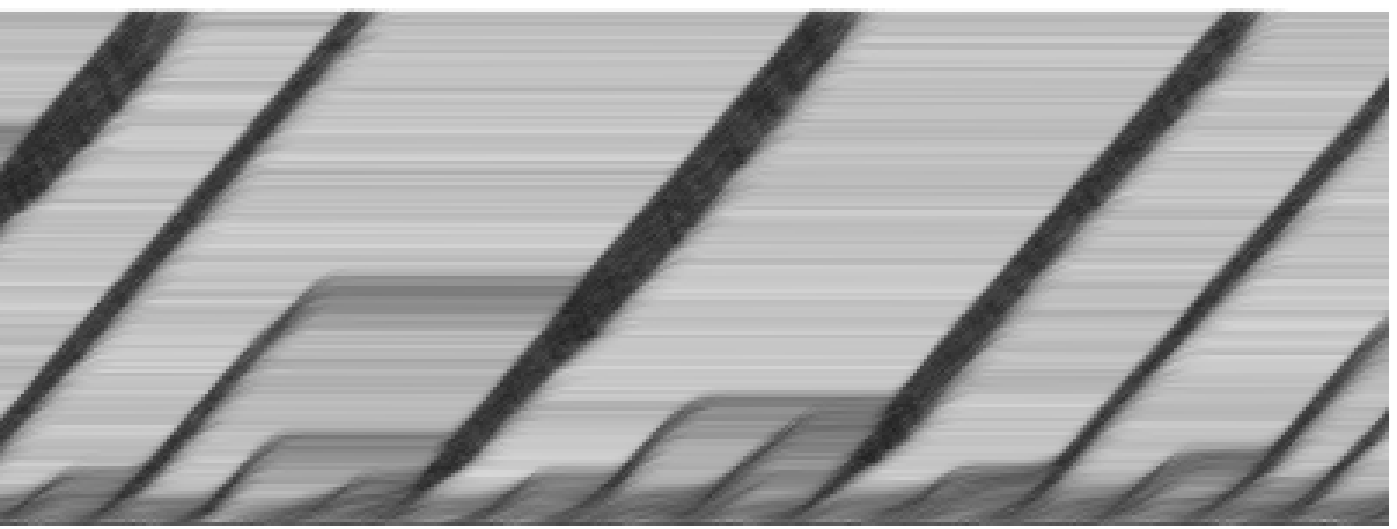}\\
\hspace*{-4ex}(b)\\[-1ex]
\includegraphics[width=1\columnwidth]{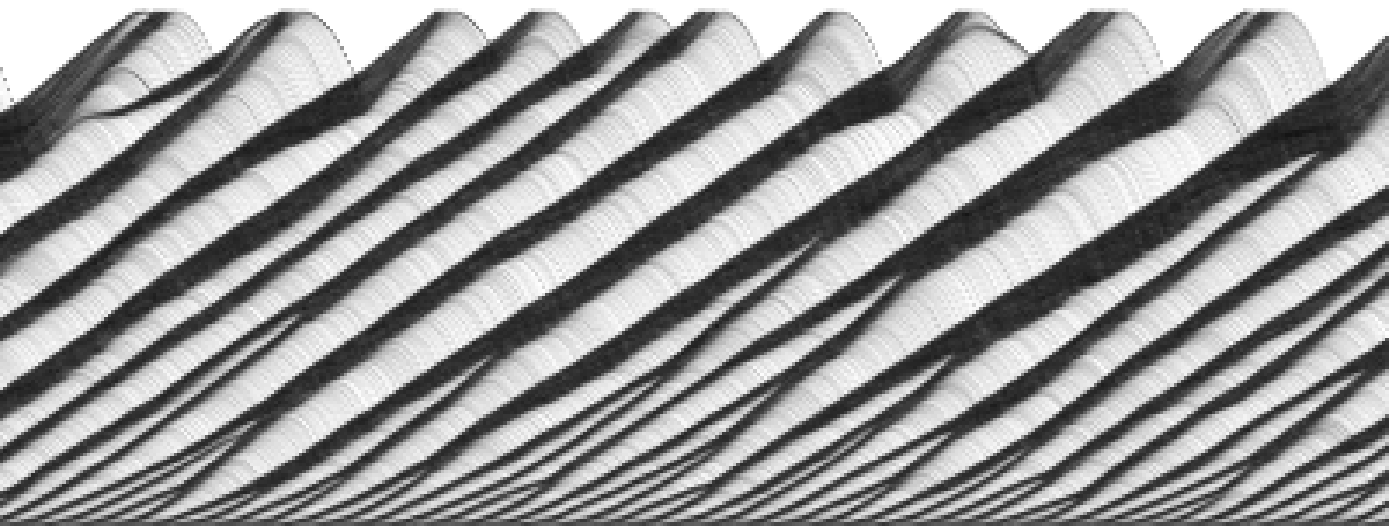}
\end{minipage}
\caption{%
Spatio-temporal images of the porosity along the tube
for (a) the horizontal and (b) the vertical plug conveying.
All the parameters, except for the direction of gravity are the same in both
simulations.
}
\label{fig:W1020_W0126}%
\end{figure}

As expected from experiments, plug conveying is observed in both the horizontal 
and the vertical cases.
However the details of the flow patterns
and the quantitative properties are different.

The most conspicuous difference between the flow patterns is
that in the horizontal transport most slugs dissolve 
while traveling along the tube.
Succeeding slugs grown through collecting the remnants of preceding slugs.
Contrary to the vertical transport, there is a final slug length.
After reaching certain length slugs do not dissolve any more.
In the vertical transport the growth of plugs comes through the merging of
smaller plugs.
Vertical plugs are slower ($0.17\,m/s$) 
then in horizontal transport ($0.3\,m/s$),
as the drag force acting on the plugs is partly 
compensated by the gravitational force.
The initial number of plugs at the beginning of the tube is lower
in the horizontal conveying, because the minimal number of particles needed to
fill up the cross-section of the tube is higher.

Another feature also observed in experiments is
that in horizontal transport the particles between 
the slugs form a resting layer at the bottom of the tube.
In vertical case the corresponding particles are accelerating
downwards, the impact of these particles on a succeeding plug is 
considerable higher then in the horizontal case.

The characteristic curves show some significant differences between the
transport modes.
Generally the pressure drop in the tube filled with plugs for
the vertical transport is about four times higher than
for the horizontal transport.
The additional pressure drop is needed to carry the plugs 
against the gravitational force.
The qualitative behavior of the pressure drop with the superficial gas velocity
and the viscosity is the same for both cases. 
The pressure drop in slug conveying decreases about six times more rapidly
with increasing superficial gas velocity.
It is decreasing less with increasing gas viscosity.
In both cases the pressure drop shows
a linear dependency on the mass flux of the granulate.
An increase of the mass flux results in more plugs along the tube.

\begin{figure}[h]
\centering
\begin{minipage}{0.85\columnwidth}
(a)\\[-1ex]
\includegraphics[width=1\columnwidth,bb=50 50 320 302]{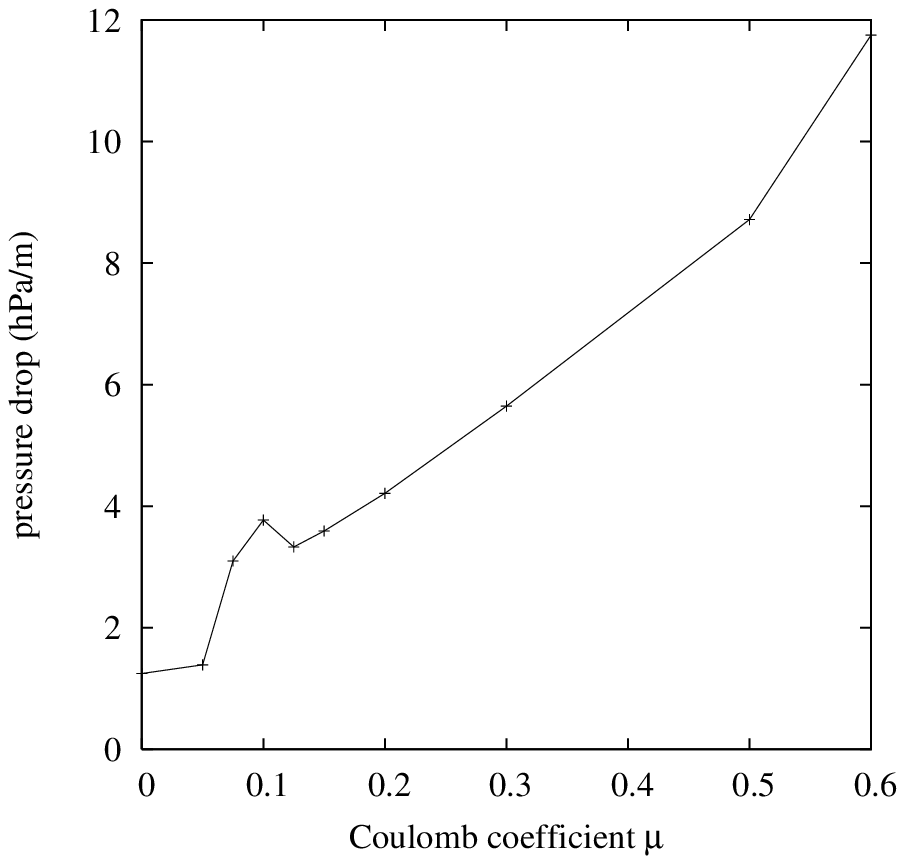}\\
(b)\\[-1ex]
\includegraphics[width=1\columnwidth,bb=50 50 320 302]{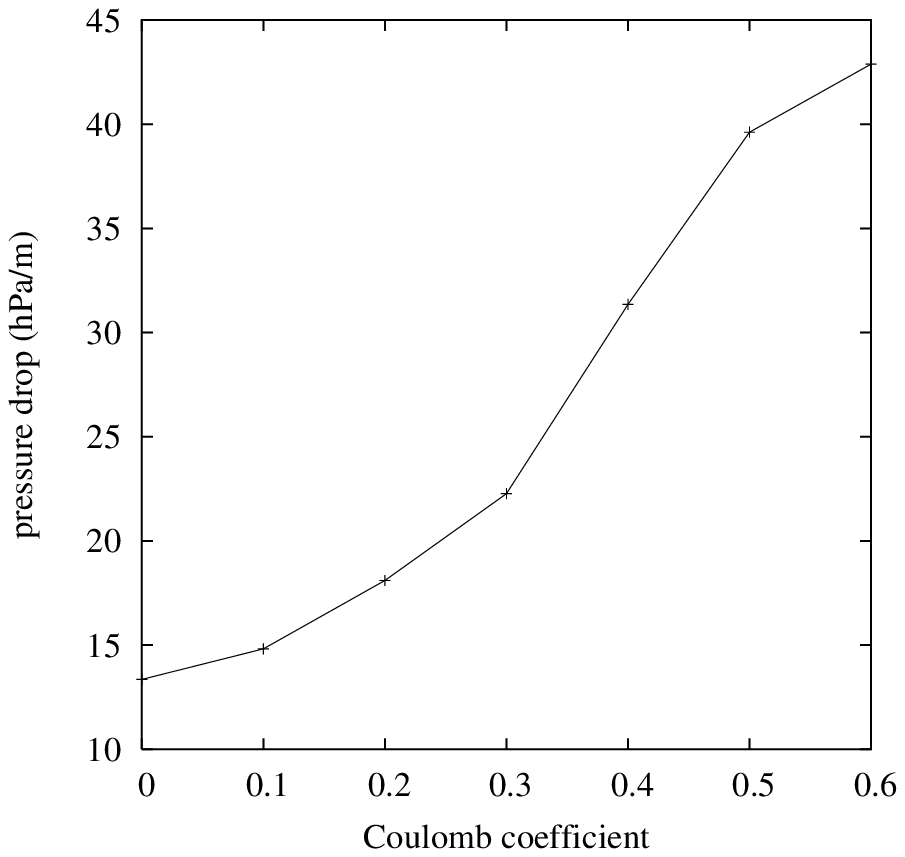}
\end{minipage}
\caption{%
Dependency of the pressure drop on the Coulomb coefficient
for (a) the horizontal and (b) the vertical plug conveying.
}
\label{fig:W1020_W0126_cc_dp}%
\end{figure}

The dependence on friction 
is different between horizontal and vertical conveying,
as can be seen from figure~\ref{fig:W1020_W0126_cc_dp}.
While in the horizontal case for low Coulomb coefficients $\mu<0.05$ 
the granular material is transported
as single particles, which are rolling on the bottom of the tube, 
in the vertical case the granular material is transported as plugs
even with no friction at all.
Here the pressure drop in the horizontal case is considerable lower
then anywhere else. 
A peak in the pressure drop ($\mu=0.1$) for the horizontal transport
denotes transition from the transport of particle layers 
at the bottom of the tube by sliding or rolling to slug transport.
On the other hand, the pressure drop increases monotonically for all values
of $\mu$ in vertical conveying.
For high Coulomb coefficients ($\mu>0.3$) in the horizontal case,
the pressure drop increases almost linearly. 
In the vertical case, the increase of the pressure drop is not linear.
As one can see from Fig 21b, there is a change in slope near $\mu=0.5$.
This change seems to be associated with the appearance of sticking plugs.

A detailed view on the plug profiles exhibits more differences.
Contrary to the vertical plugs, horizontal slugs have smooth boundaries
due to the slopes at the ends of the slug.
Within the vertical plug a slightly lower porosity is reached
(vert.\ 0.47, horiz.\ 0.49).
The slugs do not have a constant granular velocity.
The average granular velocity along the horizontal slug increases 
from both sides until the middle of the slug.
Also a shearing of particle layers within the slug is observed,
where the upper particles are the fastest.
The granular velocity is aligned in the direction of transport
all along the tube, which is not true
for the particles between the plugs in the vertical case.
In the vertical case only a small difference in the granular velocity
of the radial particle layers is observed, within the plug these
layers have a constant velocity.

The magnitude of the maximal wall stress within a plug 
is of the same order for horizontal and vertical transport.
The increase and decrease of the stress is spread over the whole plug length.
The biggest difference in the examined plug profiles is found for the 
granular temperature.
While in the vertical transport 
high granular temperatures are limited to the upper front of the plugs
and vanish rapidly within the plug due to damping,
in the horizontal transport the temperature decreases linearly along
the slug and reaches zero far behind the slug (fig.~\ref{fig:eneryyz_x}).

%
%
%
%
\section{Conclusion}
In this paper, a simple model~\cite{MCN0003} with coupled grain and gas flow
has been applied to pneumatic transport.
The implementation is three-dimensional; rotation
and Coulomb friction are taken into account.
The fluxes of gas and grains are set by the boundary conditions.
Slug conveying is observed. 
The simulation used for slug statistics and profiles contained
on average 3200 particles in a tube of length $0.525\,m$ 
and diameter $7\,mm$.
We simulated $22.5\,s$;
during this time 27 different slugs were observed
at the middle of tube, each with about 500 particles.
Additionally 24 simulations were performed to obtain the 
characteristic curves.

As for the vertical plug conveying \cite{STR0192},
an effective viscosity $\eta$ and an effective friction $\mu$
has been introduced. 
The effective viscosity reflects the increased momentum transport
in the gas due to the turbulent flow around the grains.
The effective friction reflects the complex interplay between sliding,
rolling and static friction.
Large numbers of slugs could be studied, and their porosity, velocity,
and size were measured as functions of height.

The simulation results imply that the slug formation
at the beginning of the tube occurs spontaneously.
There is a well defined preferred velocity of the slugs,
which is independent of the slug size and the tube length.
The average slug length increases along the tube due to the collection
of particles resting between the slugs.
The results also show that boundary effects at the bottom and the top
of the tube have to be taken into account. 
For the experimental parameters, the influence of the boundaries is limited 
to a distance of $0.05\,m$,
which is about the size of a slug in the middle of the tube. 
Slugs accelerate in these boundary regions.
The acceleration at the upper boundary arises due to the lack of
resting grains before the slug.
This implies that the momentum transfer to accelerate these grains can not be neglected.
A model of slug conveying must take this into account, as 
these grains reduce the slug velocity.

Also a detailed view of an average slug was presented.
In contrast to experimental setups, we are not limited to measure the 
slug properties only at few locations.
Experimental results usually provide the slug properties as a function of
time at a given position along the tube, with the disadvantage that the
slug profile is distorted by the relative motion of granulate along
the slug. 
The vertical profiles were given for porosity, granular velocity,
interparticle and drag forces, wall stress and granular temperature.
In the experiment, these parameters usually are not accessible along the whole tube.

Additionally cross-section profiles are given at high resolution.
The results show that the grains are ordered into horizontal layers.
These layers have different velocities with a fluctuation 3.5 times
larger than the mean value. 
A shearing of the particle layers is observed, where
the lowest layers are the slowest and 
the highest layers are the fastest.

A comparison of the horizontal slug conveying with the vertical conveying
shows that the flow patterns, the characteristic curves, and the plugs differ.
While in vertical conveying small plugs typically merge to bigger plugs,
the small slugs in horizontal conveying typically dissolve.
In the vertical case plugs are influenced by the distance to preceding plugs,
which is not true for the horizontal conveying.
Generally the total pressure drop in the vertical case is about four times
higher as in the horizontal case.
In horizontal conveying an additional conveying mode is observed
for low friction,
where particles slide or roll at the bottom of the tube.
In vertical conveying a sticking of plugs within the tube is found
for high friction or low superficial velocities.
The radial symmetry of the plug is broken for horizontal transport,
which comes along with a shearing of vertical particle layers.

In conclusion, the model presented here is a useful tool for investigating
slug conveying.
It is fast and flexible enough to make parameter studies 
on full featured slug conveying
and provides at the same time access to the slug properties at the level
of grains.
It can obtain both characteristic curves and slug profiles.
Future optimizations and faster computers 
will permit this model to be applied to
industrial-sized systems.
As a next step, angular periodic boundary conditions
will be applied to vertical transport.
In this way, simulations at higher tube diameters will be possible.
%
%

\begin{acknowledgements}
We thank Ludvig Vinningland and Eirik Flekk{\o}y for much help in the
fluid solver.
This research was supported by DFG (German Research Community)
contract HE 2732/2-1 and HE 2732/2-3.
\end{acknowledgements}

%
\bibliographystyle{spbasic}
\bibliography{icp-paper.bib}   
\end{document}